\documentclass[onecolumn,showpacs,preprintnumbers,amsmath,amssymb]{revtex4}
\usepackage{graphicx}
\usepackage{dcolumn}
\usepackage{bm}

\usepackage{epsfig}
\newcommand{\be}{\begin{equation}}
\newcommand{\ee}{\vspace{0cm} \end{equation}}
\newcommand{\beqq}{\setlength\arraycolsep{2pt}\begin{eqnarray}}
\newcommand{\eeqq}{\vspace{0cm} \end{eqnarray}}

\def\ie{{\it i.e.},~}
\def\eg{{\it e.g.},~}

\newcommand\la{\lower0.6ex\vbox{\hbox{\ensuremath{\buildrel{\textstyle<}\over{\sim}\ }}}}
\newcommand\ga{\lower0.6ex\vbox{\hbox{\ensuremath{\buildrel{\textstyle>}\over{\sim}\ }}}}

\newcommand{\omb}{\ensuremath{\Omega_{\rm B}\;}}

\begin{document}

\title{An Accelerating Cosmology Without Dark Energy}

\author{G. Steigman$^{1}$}\email{steigman@mps.ohio-state.edu}

\author{R. C. Santos$^{2}$}\email{cliviars@astro.iag.usp.br}

\author{J. A. S. Lima$^{3}$}\email{limajas@astro.iag.usp.br}

\affiliation{Physics Department, The Ohio State University\\  
191 West Woodruff Avenue, Columbus, OH 43210, USA}

\affiliation{Departamento de Astronomia, Universidade de S\~ao Paulo\\
Rua do Mat\~ao, 1226, 05508-900, S\~ao Paulo, SP, Brazil}

\begin{abstract}

The negative pressure accompanying gravitationally-induced particle
creation can lead to a cold dark matter (CDM) dominated, accelerating 
Universe (Lima et al.~1996~\cite{lima96}) without requiring the presence 
of dark energy or a cosmological constant.  In a recent study Lima et 
al.~2008 \cite{lss} (LSS) demonstrated that particle creation driven 
cosmological models are capable of accounting for the SNIa observations 
\cite{sn1a} of the recent transition from a decelerating to an accelerating 
Universe.  Here we test the evolution of such models at high redshift 
using the constraint on $z_{eq}$, the redshift of the epoch of matter 
-- radiation equality, provided by the WMAP constraints on the  early 
Integrated Sachs-Wolfe effect (ISW)~\cite{komatsu}.  Since the contribution 
of baryons and radiation was ignored in the work of LSS, we include 
them in our study of this class of models.  The parameters of these 
more realistic models with continuous creation of CDM is tested and 
constrained at widely-separated epochs ($z \approx z_{eq}$ and $z 
\approx 0$) in the evolution of the Universe.  This comparison reveals 
a tension between the high redshift CMB constraint on $z_{eq}$ and 
that which follows from the low redshift SNIa data, challenging the 
viability of this class of models. 

\end{abstract}


\maketitle

\section{Introduction}

A large amount of complementary cosmological data have established the surprising
result that the Universe experienced a recent transition from a decelerating to an accelerating expansion~\cite{accel}.  
The current data are consistent with a flat, Friedman-Lemaitre cosmology whose dominant components at present consist of matter 
(cold dark matter plus baryons) and a cosmological constant ($\Lambda$) or, 
equivalently, the energy density of the vacuum; the $\Lambda$CDM concordance 
model~\cite{darkenergy}.  The matter and radiation dominate earlier in the evolution 
of the Universe, leading to its decelerating expansion, while the cosmological 
constant or vacuum energy component has come to dominate more recently, driving 
the accelerating expansion.  While the concordance model provides the simplest, 
most economical explanation which is consistent with all extant data, provided that 
it is fine-tuned to fit that data, it is by no means the only candidate proposed to 
explain the accelerating expansion of the current Universe.  Dozens (if not hundreds!) 
of alternate possibilities have been explored in the literature (see Lima et al.~2008 \cite{lss} (LSS) for an extensive list of references) and many of them remain viable candidate models.  Since the astronomical community is planning a variety of large observational projects intended to test and constrain the standard $\Lambda$CDM 
concordance model, as well as many of the proposed alternative models, it is timely 
and important to identify and explore a variety of physical mechanisms (or substances) 
which could also be responsible for the late-time acceleration of the Universe. 

In principle, any homogeneously distributed ``exotic" fluid endowed with a sufficiently negative pressure, $p < -\rho/3$, so-called dark energy, will suffice. The simplest candidates for the dark energy are a positive cosmological constant $\Lambda$, or a 
non-zero vacuum energy density.  In both cases the equation of state of the exotic 
fluid is $w \equiv p/\rho = -1$.  All current observational data are in good agreement 
with the cosmic concordance ($\Lambda$CDM) model, consisting of a cosmological constant 
or vacuum energy ($w_{\Lambda} = -1$) whose current density parameter is $\Omega_{\Lambda} \approx 0.74$, plus cold dark matter and baryons ($w_{\rm M} = 0$) with $\Omega_{\rm M} 
= \Omega_{\rm CDM} + \Omega_{\rm B} = 1 - \Omega_{\Lambda} \approx 0.26$. Nevertheless, 
the $\Lambda$CDM model encounters several challenges.  For example, the value of the 
energy density of the vacuum required by the data ($\Omega_{\Lambda} \approx 0.26$)
needs to be fine-tuned, exceeding the estimates from quantum field theories by some 
50 -- 120 orders of magnitude.  If the vacuum energy (or $\Lambda$) differs from zero, 
why does it have the value needed for consistency with the data?

LSS noted recently \cite{lss} that the key ingredient required to accelerate the 
expansion of the Universe is a sufficiently negative pressure, which occurs naturally 
as a consequence of cosmological particle creation driven by the gravitational field
\cite{partcreat}.  While completely different from it, there are some analogies between 
this class of models and the continuously accelerating, never decelerating, expansion 
of the steady-state cosmology driven, in the Hoyle-Narlikar \cite{hoyle} version, by 
C-field induced particle creation.  The late-time evolution of the class of models 
explored by LSS is qualitatively very different from that of the standard, $\Lambda$CDM concordance model in that they lack a cosmological constant and, when baryons are 
included, the ratio of the dark matter to baryon densities evolves with time or 
redshift.  Nonetheless, the LSS \cite{lss} analysis of the late-time evolution of 
such a CDM-dominated model with no cosmological constant or dark energy, lacking 
baryons and radiation, showed that it is quantitatively possible to account for 
the SNIa data which reveal a recent transition from decelerated to accelerated 
expansion.  Indeed, while the flat $\Lambda$CDM concordance model provides a good 
fit to the SNIa data, the best fit actually corresponds to a positively curved 
(closed) $\Lambda$CDM model, so that for some choice of parameters the models 
considered by LSS may provide an even better fit to the supernovae data.  For 
a subset of the models explored by LSS to be discussed here, the early-time, 
high redshift (\eg $z~\ga 10$) evolution is indistinguishable from that of 
the $\Lambda$CDM concordance model, so that the parameters controlling the 
early evolution of these models must agree with those of the concordance model.  
To set the stage for the constraints and tests discussed here, we note that for 
the $\Lambda$CDM concordance model with $\Omega_{\Lambda} = 1 - \Omega_{\rm M} = 
0.76$, the redshift of the transition from decelerated to accelerated expansion 
is $z_{t} = 0.79$, the redshift of the epoch of equal matter and radiation 
densities (related to the early, Integrated Sachs-Wolfe (ISW) effect) is 
$z_{eq} = 3223$ and, the present age of the Universe in units of the Hubble 
age ($H_{0}^{-1}$) is $H_{0}t_{0} = 1.00$. 

Since at present baryons are subdominant to CDM and the contribution of radiation is negligible, the neglect of baryons and radiation by LSS was not a bad approximation for 
their study of the recent evolution of the Universe.  Here, we revisit the LSS model 
of dark matter creation driven acceleration in a more realistic model incorporating 
baryons and radiation.  While it will be seen that there is a subset of models with 
baryons and radiation whose late time evolution (\ie $H = H(z)$ for $z~\la 5$) is sufficiently close to those studied by LSS  and to the $\Lambda$CDM concordance model 
(and, therefore, they too will be consistent with the SNIa data), these models differ 
from those studied by LSS in their early evolution, in particular at the epochs of recombination and of equal matter and radiation densities.  To further test these 
models we impose an additional, high-redshift constraint derived from the early 
Integrated Sachs-Wolfe (ISW) effect, provided by the WMAP~\cite{komatsu} value 
of the redshift of matter -- radiation equality, $z_{eq} = 3141 \pm 157$.  

In \S~\ref{partcreat} we describe the class of models under consideration and solve 
for the evolution of the Hubble parameter as a function of redshift ($H = H(z)$).  
In \S~III we compare the predictions of this class of models as a function of the 
two parameters related to the particle creation rate to the SNIa data~\cite{sn1a} 
and to the constraint on the redshift of the epoch of equal matter and radiation 
densities provided by the WMAP observations of the CMB temperature anisotropy spectrum~\cite{komatsu}.  Our results are summarized and our conclusions presented 
in \S~\ref{concl}.

\section{Particle Creation Driven Cosmology}
\label{partcreat}

Consider a sufficently large comoving volume ($V$), representative on average of 
the Universe which, at some time, contains $N = nV$ CDM particles ($w_{\rm CDM} 
= 0$).  If CDM particles are being created at the expense of the gravitational 
field \cite{partcreat}, then
\be
{1 \over N}{dN \over dt} = \Gamma,
\ee
where $\Gamma$ is the creation rate (number of particles per unit time), assumed 
to be uniform throughout the Universe.  In this case, since the mass density in 
the CDM is proportional to the number density, 
\be
{d[{\rm ln}(\rho_{\rm CDM}V)] \over dt} = \Gamma,
\ee
so that
\be
{\rho_{\rm CDM} = \rho_{{\rm CDM},0}(1 + z)^{3}{\rm exp}[-\int^{t_{0}}_{t}\Gamma dt'}].
\ee
In eq.~3, $z$ is the redshift ($V = V_{0}(1 + z)^{-3}$) and the subscript 0 is for 
quantities evaluated at the present epoch ($z = 0$).

The models we consider are defined by the choice for the particle creation rate, 
$\Gamma$.  The most natural choice would be a particle creation rate which favors 
no epoch in the evolution of the Universe, such as $\Gamma = 3\beta H$, where $H$ 
is the Hubble parameter and the free parameter $\beta$ is  positive.  However, in the absence of baryons and radiation, it is 
easy to show (see LSS) that such a Universe will always decelerate if $\beta < 1/3$ 
and will always accelerate if $\beta > 1/3$.  The inclusion of baryons and radiation, 
whose effect is to drive a decelerated expansion, opens the possibility of a 
transition from an early, matter-dominated deceleration, to a late, CDM-driven 
accleration.  In this case the transition from decelerating to accelerating 
expansion is set by the baryon density and it occurs ``naturally", late in the 
evolution of the Universe at low redshift, without any need for fine-tuning.   
However, it will be seen that the early evolution of models with this form for 
the particle creation rate ($\Gamma \propto H$) are inconsistent with the early 
ISW data.  This problem may be alleviated by adding a second creation term for 
the particle creation rate which is ``tuned" to the current epoch, $\Gamma 
\propto H_{0}$, where $H_{0}$ is the present ($t_{0}$) value of the Hubble 
parameter.  As a result, the class of particle-creation driven models to 
be studied here are defined by,
\be
\Gamma = 3\gamma H_{0} + 3\beta H,
\ee
where $0 \leq \{\gamma,\beta\} \leq 1$.  In this case, the CDM mass/energy density
evolves as,
\be
\rho_{\rm CDM} = \rho_{{\rm CDM},0}(1 + z)^{3(1 - \beta)}{\rm exp}[3\gamma(\tau 
- \tau_{0})],
\ee
where $\tau \equiv H_{0}t$ and $\tau_{0} \equiv H_{0}t_{0}$ are the age and the 
present age of the Universe in units of the Hubble age ($H_{0}^{-1}$).  For $H_{0}
\equiv 100h$~kms$^{-1}$Mpc$^{-1}$, $H_{0}^{-1} = 9.78h^{-1}$~Gyr.  In our analysis
the HST Key project result~\cite{hst} $h = 0.72 \pm 0.08$ is adopted, so that
$H_{0}^{-1} = 13.6 \pm 1.5$~Gyr.

For simplicity, as is the case for the cosmic concordance model, our considerations 
are limited to flat cosmologies, so that $\Omega_{\rm M} = \Omega_{\rm B} + 
\Omega_{\rm CDM} = 1$, neglecting the very small contribution from the radiation 
density at present, when $\Omega_{\rm B}$ and $\Omega_{\rm CDM}$ are evaluated.  
For flat models, the general Friedman equation reads,
\be
\left({H \over H_{0}}\right)^{2} = \Omega_{\rm R}(1 + z)^{4} + \Omega_{\rm B}(1 + z)^{3} 
+ (1 - \Omega_{\rm B})(1 + z)^{3(1 - \beta)}{\rm exp}[3\gamma(\tau - \tau_{0})].
\ee
In comparing the model predictions with the SNIa data, the radiation density term
($\Omega_{\rm R} < 10^{-4}$) may be safely neglected.  

At the time of Big Bang Nucleosynthesis (BBN) the Universe is radiation dominated 
and only the baryon density (along with the radiation density) plays an important 
role.  BBN cares about neither the dark matter or the cosmological constant for the 
$\Lambda$CDM model, nor about the creation of dark matter for the \{$\beta, \gamma$\}
model considered here.  However, since the creation of dark matter is accompanied by 
a ``creation pressure" ${\it p}_{c} = -\Gamma\rho_{\rm DM}/3H$ (see LSS), and the 
evolution of the dark matter density differs from the ``usual" $(1 + z)^{3}$ evolution 
for conserved particles, the late time growth of perturbations in this model will 
likely depart from that in the concordance model.  At very early times near the 
epochs of equal matter and radiation densities and recombination, $w_{c} \equiv 
{\it p}_{c}/\rho_{\rm DM} \rightarrow -\beta$.  So if, as will be seen below, 
$\beta = 0$ or $\beta \ll 1$ are favored, the early growth of perturbations in 
the \{$\beta, \gamma$\} model will track that of the concordance model.  As a 
result, for the second observational constraint we require that the model-predicted 
redshift of the epoch of equal matter and radiation densities, $1 + z_{eq} \equiv 
\rho_{\rm M}/\rho_{\rm R}$, agree with that determined by the WMAP observations 
of the early ISW effect~\cite{komatsu}.  Since the $\Lambda$CDM model is consistent 
with these observations, the early-time ($t \ll t_{0}$) 
evolution of these $\{\beta, \gamma\}$ models must track closely that of the standard $\Lambda$CDM model.  

Before considering the general case, where $\{\Omega_{\rm B}, \beta, \gamma \} 
\neq 0$, it is instructive to examine several simplified cases where one or more 
of these parameters is set equal to zero.

\subsection{The LSS Model: $\Omega_{\rm B} = 0$}

Ignoring the contributions from baryons and radiation, this is the model 
studied by LSS~\cite{lss}.  In this case the Friedman equation simplifies to,
\be
\left({H \over H_{0}}\right)^{2} = \left[{1 \over a}\left({da \over 
d\tau}\right)\right]^{2} = (1 + z)^{3(1 - \beta)}{\rm exp}[3\gamma(\tau 
- \tau_{0})],
\ee
which has a solution for the scale factor $a(t)$ (or the redshift, $z$) as a 
function of time, 
\be
{a(t) \over a_{0}} = {1 \over (1 + z)} = \left[\left({1 - \gamma - \beta \over 
\gamma}\right) ({\rm e}^{3\gamma\tau/2} - 1)\right]^{{2 \over 3(1 - \beta)}}.
\ee
Evaluating this at $z = 0$, where $\tau = \tau_{0}$ and $a = a_{0}$, relates 
the present age, $\tau_{0}$, to the \{$\beta, \gamma$\} parameters,
\be
{\rm exp}(3\gamma\tau_{0}/2) = {1 - \beta \over 1 - \gamma - \beta}.
\ee
The present age of the Universe ($t = t_{0}, \tau = \tau_{0}$) is
\be 
\tau_{0} = H_{0}t_{0} = {2 \over 3\gamma}{\rm ln}({1 - \beta \over 
1 - \gamma - \beta}).
\ee
Solving for the time -- redshift relation
\be
{\rm exp}(3\gamma\tau/2) = 1 + \left({\gamma \over 
1 - \gamma - \beta}\right)(1 + z)^{-3(1 - \beta)/2}.
\ee
Using the $z$ versus $\tau$ and $\tau_{0}$ relations above, the 
Friedman equation for $H = H(z)$, reduces to the simple form
\be
H/H_{0} = (1 - \beta)^{-1}[\gamma + (1 - \gamma - \beta)(1 + z)^{3(1 - \beta)/2}].
\ee

It is easy to confirm (see LSS for the details) that if $\gamma = 0$, the expansion of 
the Universe always decelerates for $0 \leq \beta < 1/3$ and always accelerates for 
$1/3 < \beta \leq 1$.  In this case there is no transition from an early decelerating 
to a late accelerating Universe.  However, with $\gamma \neq 0$ ($0 \leq \gamma \leq 1$) 
the redshift, $z_{t}$, of the transition from early-time deceleration to late-time acceleration is given by
\be
1 + z_{t} = [{2\gamma \over (1 - 3\beta)(1 - \gamma - \beta)}]^{{2 \over 3(1- \beta)}}.
\ee

\begin{figure}[!ht]
\begin{center}
\epsfig{file=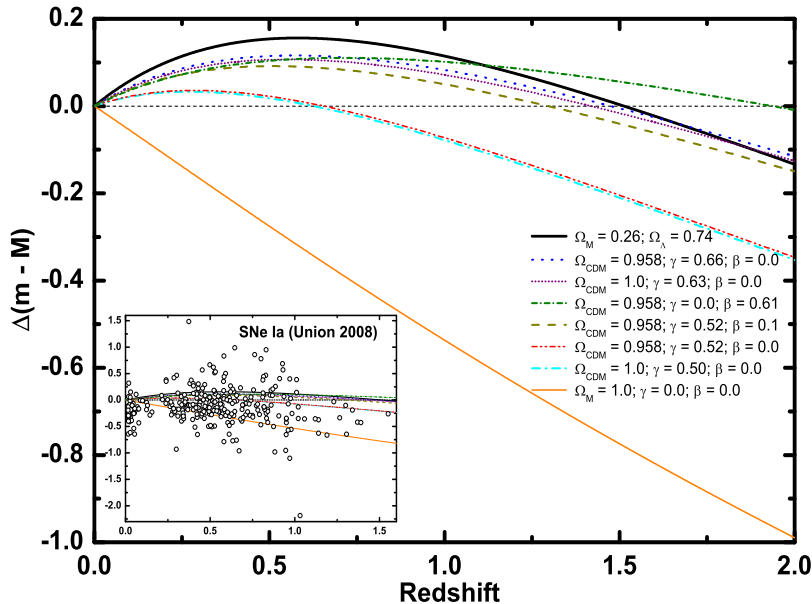, width=4.48truein,height=3.5truein}
\caption{The relative distance modulus ($\Delta (m - M)$) -- redshift
relations for a variety of models with and without particle creation.  
For models without particle creation the $\Lambda$CDM model (solid, 
black) and the Einstein-deSitter model (solid, orange) are shown.  
Several particle creation models are shown, with and without 
baryons, for a selection of \{$\beta, \gamma$\} choices. The inset 
figure shows the same curves along with the central values from the 
SNIa data~\cite{sn1a}.}
\label{fig:distmod}
\end{center}
\end{figure}

\begin{figure}[!ht]
\begin{center}
\epsfig{file=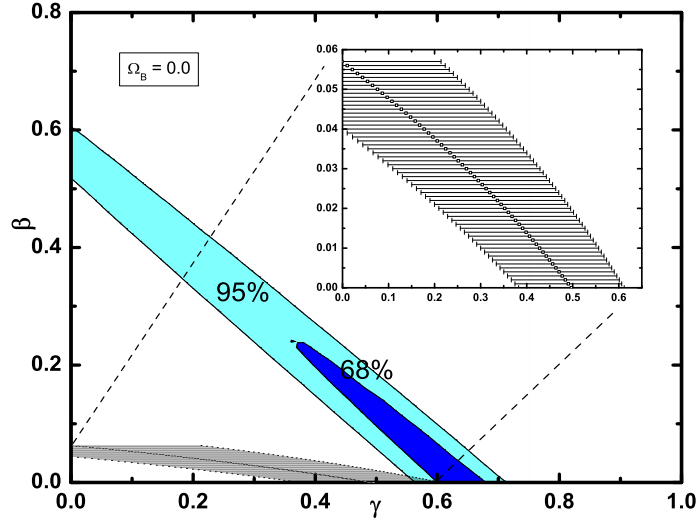, width=4.0truein,height=3.125truein}
\caption{The 68\% (dark blue) and 95\% (light blue) contours in 
the $\beta - \gamma$ plane derived for \omb = 0 from the SNIa 
data~\cite{sn1a}, along with the 95\% confidence band (black; see 
the inset which zooms in on this band) consistent with the early 
ISW constraint that $z_{eq} = 3141 \pm 157$~\cite{komatsu}. The 
SNIa data are best fit at $\beta = 0$ and $\gamma = 0.63$.  The 
solid black curve, shown in more detail in the inset, is for the 
best fit relation between $\beta$ and $\gamma \equiv \gamma^{0}$ 
(see eqs. 16,17) consistent with the observed value of $z_{eq}$ 
\cite{komatsu}.} 
\label{fig:betagamma0}
\end{center}
\end{figure}

\begin{figure}[!ht]
\begin{center}
\epsfig{file=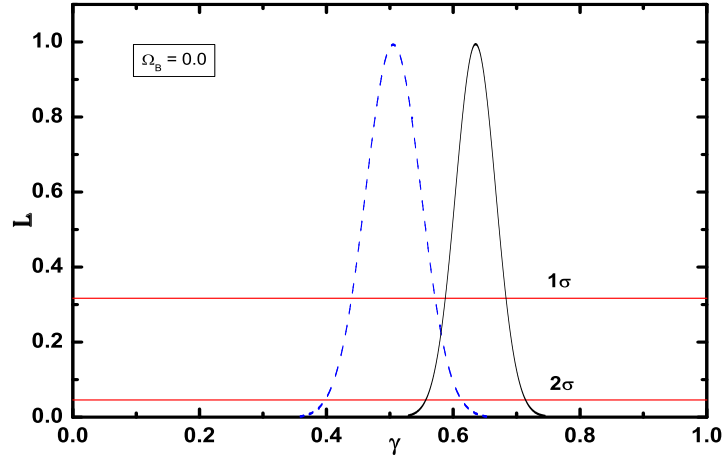, width=4.0truein,height=3.125truein}
\caption{The likelihood functions for $\gamma \equiv \gamma^{0}$ 
inferred from the SNIa data (solid/gray) and from the $z_{eq}$ 
constraint (dashed/blue) for $\beta = 0$ and \omb = 0.} 
\label{fig:probgamma2}
\end{center}
\end{figure}

As shown in LSS, there is a range of \{$\gamma, \beta$\} values whose fit to 
the SNIa data is at least as good as the fit provided by the parameters of the 
$\Lambda$CDM concordance model.  In Figure~\ref{fig:distmod} we illustrate this 
by comparing the relative distance modulus -- redshift relations for several 
combinations of $\{\gamma, \beta \}$ with that for the $\Lambda$CDM model and, 
for comparison with a model which leads to a poor fit, the Einstein -- deSitter 
model.  Note that for $z ~\la 1.5$, many of the \{$\beta, \gamma$\} models, with 
and without baryons, are indistinguishable from the $\Lambda$CDM model.  
While the curves for the best fit values for $\beta = 0$ 
and \{$\Omega_{\rm B}, \gamma$\} = \{0, 0.63\} and \{0.042, 0.66\} are very close to each other, the 
curve for $\gamma = 0$, \omb = 0.042 and $\beta = 0.61$ deviates noticeably from
them, especially at $z~\ga 1$.  Furthermore, the curves for $\beta = 0$ and 
$\gamma = 0.52~(0.50)$, corresponding to the $z_{eq}$ constraint with (without)
baryons, deviates significantly from those curves which provide the best fits
to the SNIa data.

For the Union 2008 set of SNIa data~\cite{sn1a} adopted here, the best fitting 
combination of parameters (for \omb = 0.042) occurs for $\beta = 0$ and $\gamma 
= 0.63$, corresponding to $z_{t} = 1.26$.  While this value for the transition 
redshift may seem surprisingly high compared to $z_{t} = 0.79$ for the $\Lambda$CDM 
concordance model with $\Omega_{\rm M} = 1 - \Omega_{\Lambda} = 0.26$, as may 
be seen from Figure \ref{fig:distmod} the relative distance modulus -- redshift 
relations for these two models are very similar at redshifts $z~\la 1.5$.  For 
$\beta = 0$ and $\gamma = 0.63$, the present age of the Universe is $\tau_{0} 
= H_{0}t_{0} = 1.05$, corresponding to $t_{0} = 14.3$~Gyr.
 
At high redshifts, including the contribution from radiation (but not yet from 
baryons), the Friedman equation is modified to
\be
\left({H \over H_{0}}\right)^{2} \approx \Omega_{\rm R}(1 + z)^{4} + 
\left({1 - \gamma - \beta \over 1 - \beta}\right)^{2}(1 + z)^{3(1 - \beta)},
\ee
so that the redshift of equal matter (CDM) and radiation densities in this 
class of models, without baryons ($\Omega_{\rm B} = 0$), is given by
\be
\Omega_{\rm R}(1 + z_{eq})^{1 + 3\beta} = \left({1 - \gamma - \beta \over 
1 - \beta}\right)^{2}.
\ee
Fixing $z_{eq}$, along with the observed value of $\Omega_{\rm R}$, provides 
a complementary constraint on the $\gamma - \beta$ relation to that from the 
SNIa data.  For $\Omega_{\rm B} = 0$, the $\gamma(\Omega_{\rm B} = 0) \equiv 
\gamma^{0} - \beta$ relation is
\be
\gamma^{0} = (1 - \beta)\{1 - [\Omega_{\rm R}(1 + z_{eq})]^{1/2}(1 + 
z_{eq})^{3\beta/2}\}.
\ee

Accounting for the still relativistic neutrinos and for $h = 0.72 \pm 0.08$, 
$\Omega_{\rm R} = 8.065(1 \pm 0.22)\times 10^{-5}$.  For the combination of 
parameters which best fits the SNIa data~\cite{sn1a}, $\beta = 0$ and $\gamma^{0} 
= 0.63$, $z_{eq} = 1661$, a redshift which is much to small for consistency 
with the WMAP data which finds $z_{eq} = 3141 \pm 157$ \cite{komatsu}.  

For $1 + z_{eq} = 3142(1 \pm 0.05)$,
\be 
\gamma^{0} = (1 - \beta)[1 - 0.503(1 \pm 0.114)(3142)^{3\beta/2}].
\ee
The 68\% and 95\% constraints when $\Omega_{\rm B} = 0$ for $\gamma^{0}$ versus 
$\beta$ derived from $z_{eq}$ are shown in Figure~\ref{fig:betagamma0}.  For 
$\beta = 0$, the WMAP measurement of the early ISW effect~\cite{komatsu} requires 
$\gamma^{0} = 0.50 \pm 0.06$.

It is easy to understand this result by comparing the high redshift evolution of 
the $\{\beta, \gamma \}$ model (for $\beta = 0$) with that of the $\Lambda$CDM model.  
Consistency between them requires that $(1 - \gamma^{0})^{2} \approx \Omega_{\rm M} 
= 0.26$, so that $\gamma^{0} \approx 0.49$, in excellent agreement with the more 
direct result from the early ISW effect. 

For each of the independent constraints from the SNIa data and the early ISW effect, 
the best fit occurs at $\beta = 0$ and, as may be seen from Figure~\ref{fig:betagamma0} 
the parameters identified by these two constraints diverge from each other as $\beta$ increases.  
In Figure~\ref{fig:probgamma2} are shown the probability distributions for $\gamma$ when $\beta = 0$ derived from the SNIa data~\cite{sn1a} and from the early 
ISW effect~\cite{komatsu}.   

The early ISW combination of parameters appears to be in some conflict with 
those identified by the SNIa data~\cite{sn1a}.  For example, for $\beta = 0$ 
and $\gamma^{0}_{\rm SNIa} = 0.63 \pm 0.04$, leads to a prediction of $z_{eq}$, 
$(1 + z_{eq})_{\rm SNIa} = 1661^{+544}_{-524}$, which is some $2.7\sigma$ away 
from the WMAP value~\cite{komatsu}.  We return to a more careful discussion of 
the tension between the SNIa and WMAP data for this class of models in the 
context of our discussion of the more realistic model including baryons.  
Nonetheless, the results presented here provide a useful background for the 
subsequent discussion of those cases for which $\Omega_{\rm B} \neq 0$.

\section{Including Baryons}

The analysis of LSS~\cite{lss} established that as an alternative to dark energy, 
a model with the creation of cold dark matter by the gravitational field is capable 
of accounting for the evolution of a Universe in which the early-time decelerated 
expansion is succeeded by a late-time accelerating phase, consistent with the SNIa 
data.  This encourages us to explore a more realistic version of the LSS model 
including baryons (and radiation).  While qualitatively different from the models 
explored by LSS, models with baryons are quantitatively very similar to them since 
for all intermediate and low redshifts $\rho_{\rm B} \ll \rho_{CDM}$. As in LSS, the SNIa data~\cite{sn1a} 
will be used to constrain the late-time evolution of this class of models.  To 
further test them and to constrain their parameters, the observational early ISW 
effect constraint on the redshift of equal matter and radiation densities inferred 
from WMAP, $1 + z_{eq} = 3142 \pm 157 = 3142 (1 \pm 0.05)$~\cite{komatsu} is used.

For a successful model be consistent with the early ISW effect requires that the 
early, high redshift evolution of these models be nearly identical with that of 
the $\Lambda$CDM concordance model.  However, this does not guarantee that the 
growth of structure in these models need be the same as in the concordance model.  
For the very early evolution of both models only baryons (for BBN) and radiation 
are important; $\rho_{\rm CDM}$ and $\rho_{\Lambda}$ may be neglected.  In particular, 
since the results of BBN depend only on the baryon and radiation densities and are independent of $\rho_{\rm CDM}$ and $\rho_{\Lambda}$, as well as of any particle 
creation, it is best to choose \omb from BBN and not, for example, from observations 
of large scale structure such as those provided by galaxy correlation functions.  
In the standard model, with three flavors of light neutrinos, the predicted 
primordial abundances depend only on the baryon to photon ratio which is directly 
related to $\Omega_{\rm B}h^{2}$.  To constrain $\Omega_{\rm B}$, deuterium is 
the baryometer of choice (see~\cite{Gary} for a recent review and further references).  
For BBN~\cite{vimal}, using the latest observations of deuterium~\cite{pettini}, 
$\Omega_{\rm B}h^{2} = 0.0218 \pm 0.0011$.  For the HST Key Project value of $H_{0}$~\cite{hst}, $\Omega_{\rm B} = 0.042 \pm 0.010$.  Since the models we 
consider are flat and contain no dark energy or a cosmological constant, 
$\Omega_{\rm CDM} = 1 - \Omega_{\rm B} = 0.958 \pm 0.010$.  These are the 
values adopted in our analysis below.      

\subsection{$\Omega_{\rm B} \neq 0, \beta \neq 0, \gamma = 0$}

As noted earlier and demonstrated in LSS, in the absence of baryons and when $\gamma
= 0$ the expansion of the Universe always decelerates for $0 \leq \beta < 1/3$ and 
always accelerates for $1/3 < \beta \leq 1$.  However, with the inclusion of baryons,
which dominate over the CDM at high redshifts, the early expansion of the Universe
always decelerates.  In this case, for a choice of $\beta$ in the range \{$1/3, 1$\}, 
the late time expansion will accelerate.  To better understand this transition,
consider the ``total equation of state", $w \equiv {\it p}_{\rm TOT}/\rho_{\rm TOT}$,
where ${\it p}_{\rm TOT} = {\it p}_{\rm B} + {\it p}_{\rm CDM} + {\it p}_{c} = 
{\it p}_{c}$, since ${\it p}_{\rm B} = {\it p}_{\rm CDM} = 0$.  For $w > -1/3$, 
the expansion decelerates, while for $w < -1/3$, it accelerates.  Including 
baryons, but ignoring the early-time contribution from radiation,
\be
w = {\it p}_{c}/[\rho_{\rm B} + \rho_{\rm CDM}].
\ee
Since, for $\gamma = 0$, ${\it p}_{c} = -\beta\rho_{\rm CDM}$, 
\be
w = -\beta\left[1 + \left({\rho_{\rm B} \over \rho_{\rm 
CDM}}\right)\right]^{-1} = -\beta\left[1 + \left({\Omega_{\rm B} 
\over 1 - \Omega_{\rm B}}\right)(1 + z)^{3\beta}\right]^{-1}.
\ee
At present, for $z = 0$, $w = -\beta(1 - \Omega_{\rm B})$, so that 
for \omb = 0.042, the expansion is accelerating if $\beta > 1/3(1 - 
\Omega_{\rm B}) = 0.348$.  In the future ($a \rightarrow \infty$, $z 
\rightarrow -1$), $w \rightarrow -\beta$ (for the $\Lambda$CDM model, 
$w \rightarrow -1$).

\begin{figure}[!ht]
\begin{center}
\epsfig{file=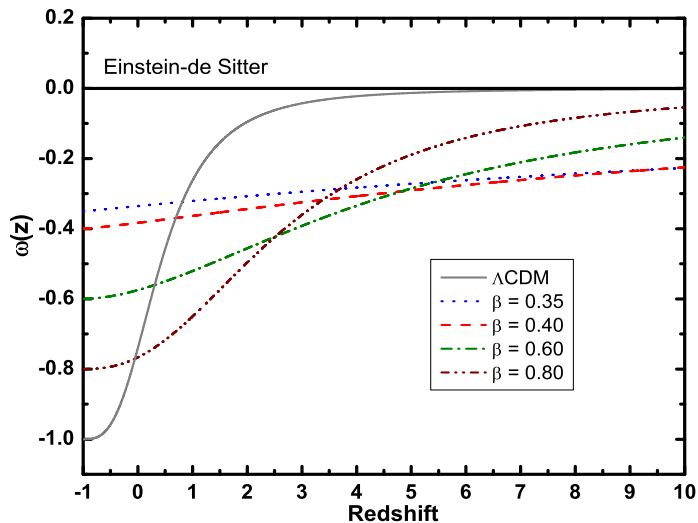, width=4.0truein,height=3.125truein}
\caption{The equation of state parameter $w$ versus redshift relations for 
\omb = 0.042 and $\gamma = 0$ for several choices of $\beta$.  Also shown 
for comparison are the $w(z)$ variations for the Einstein-deSitter model 
($w = 0$) and for the $\Lambda$CDM model (solid curves).  Notice that $z 
< 0$ corresponds to the future ($a > a_{0}$) evolution of the Universe; 
as $z \rightarrow -1$, $a \rightarrow \infty$.} 
\label{fig:wzbeta}
\end{center}
\end{figure}

\begin{figure}[!ht]
\begin{center}
\epsfig{file=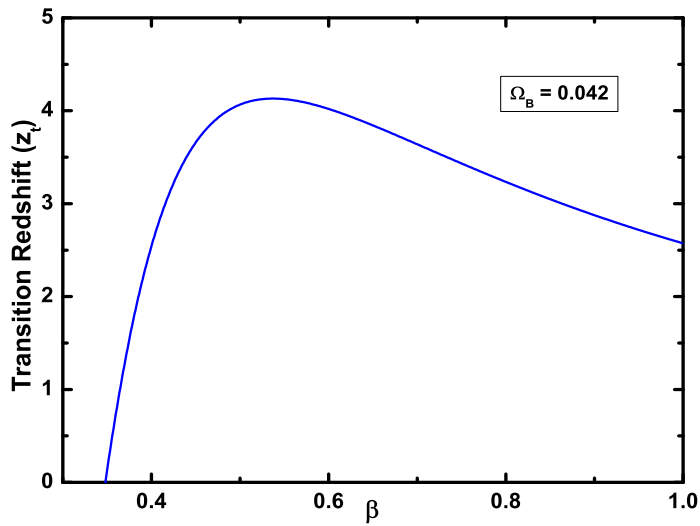, width=4.0truein,height=3.125truein}
\caption{The transition redshift, $z_{t}$, as a function of $\beta$ 
for $\gamma = 0$ and \omb = 0.042.} 
\label{fig:betazt}
\end{center}
\end{figure}

As may be seen from eq.~19, for $\gamma = 0$ and $\Omega_{\rm B} \neq 0$ the 
transition from deceleration to acceleration ($w = -1/3$) occurs at redshift 
$z_{t}$ where,
\be
1 + z_{t} = \left[\left({1 - \Omega_{\rm B} \over \Omega_{\rm B}}\right)(3\beta - 1)\right]^{1/3\beta}.
\ee
The $z_{t} - \beta$ relation for $\gamma = 0$ and \omb = 0.042 is shown in Figure \ref{fig:betazt}.  Without any fine-tuning the expansion of the Universe has a 
late-time, low-redshift ($0~\la z_{t}~\la 4$) transition from decelerating to 
accelerating.  The present epoch is ``selected" in this model by the BBN-determined 
value of $\Omega_{\rm B} \neq 0$.

In this case ($\gamma = 0$), as may be seen from Figure~\ref{fig:betagamma042}, 
the best (albeit not very good) fit (see Fig.~\ref{fig:distmod}) to the SNIa 
data~\cite{sn1a} is for $\beta = 0.61$, so that $z_{t} = 4.0$.  For this model 
the baryons dominate over the CDM during the early evolution of the Universe, 
when $z > z_{*}$, where $\rho_{\rm B}(z_{*}) \equiv \rho_{\rm CDM}(z_{*})$,
\be
1 + z_{*} \equiv \left({1 - \Omega_{\rm B} \over \Omega_{\rm B}}\right)^{1/3\beta}.
\ee
For the model with $\beta = 0.61$ and \omb = 0.042 (and $\gamma = 0$), $z_{*} 
= 4.5$.  At high redshifts, $z \gg 5$, $\rho_{\rm M} \rightarrow \rho_{\rm B}$, 
so that $1 + z_{eq} \rightarrow \Omega_{\rm B}/\Omega_{\rm R} = 522$, clearly 
in conflict with the constraint on $z_{eq}$ from the early ISW effect.  So, 
although it is of some interest that the late time evolution of this model 
with $\gamma = 0$ and $\Omega_{\rm B} \neq 0$ is capable of a transition 
from decelerated to accelerated expansion without any fine-tuning, the 
early evolution of this model is in conflict with the WMAP data.

\subsection{$\Omega_{\rm B} \neq 0, \gamma \neq 0, \beta = 0$}

Motivated by the LSS result for \omb = 0 that the best fit to the SNIa data occurs 
for $\beta = 0$, we next explore this case with \omb = 0.042.  Setting $\beta = 0$ 
but allowing for a non-zero value of the baryon density, the Friedman equation, 
including radiation, becomes
\be
\left({H \over H_{0}}\right)^{2} = \Omega_{\rm R}(1 + z)^{4} + 
(1 + z)^{3}\{\Omega_{\rm B} + (1 - \Omega_{\rm B}){\rm exp}[3\gamma(\tau - \tau_{0})]\}.
\ee 

For the late-time evolution of $H = H(z)$ needed for comparison with the SNIa data, 
the contribution from the radiation term may be neglected.  Since $\Omega_{\rm B} 
\ll 1 - \Omega_{\rm B}$ (and, it can be checked that $\Omega_{\rm B} \ll (1 - 
\Omega_{\rm B}){\rm e}^{-3\gamma\tau_{0}}$), it is a good approximation in solving 
for the $z$ versus $t$ relation to neglect the $\Omega_{\rm B}$ term in eq.~19.  If 
so, then at late-times when $z$ is not too large (and $\tau$ is not $\ll \tau_{0}$),
\be
\left({H \over H_{0}}\right)^{2} = \left[{1 \over a}\left({da \over 
d\tau}\right)\right]^{2} \approx (1 - \Omega_{\rm B})(1 + z)^{3}{\rm 
exp}[3\gamma(\tau - \tau_{0})].
\ee 
In this approximation, accounting for the $(1 - \Omega_{\rm B})$ prefactor, eq.~21 
is identical in form to eq.~7 for $\Omega_{\rm B} = 0$, so that the redshift -- age 
relation is well approximated by
\be
a(t)/a_{0} = (1 + z)^{-1} \approx (1 - \Omega_{\rm B})^{1/3}\gamma^{-2/3}
{\rm e}^{-\gamma\tau_{0}}({\rm e}^{3\gamma\tau/2} - 1)^{2/3}.
\ee
Evaluating this expression at $t = t_{0}$ ($z = 0$) establishes, in this 
approximation, the connection among $\tau_{0}$, $\gamma$, and $(1 - \Omega_{\rm B})$,
\be
{\rm exp}(-3\gamma\tau_{0}/2) \approx 1 - \gamma(1 - \Omega_{\rm B})^{-1/2}
\equiv 1 - \gamma',
\ee
where $\gamma' \equiv \gamma(1 - \Omega_{\rm B})^{-1/2}$.  The present age 
of the Universe ($t = t_{0}, \tau = \tau_{0}$) is
\be 
\tau_{0} = H_{0}t_{0} = {2 \over 3\gamma}{\rm ln}\left({1 \over 1 - \gamma'}\right).
\ee
Solving for the age -- redshift relation,
\be
{\rm exp}(3\gamma\tau/2) \approx 1 + \left({\gamma' \over 1 - \gamma'}\right)(1 + z)^{-3/2}.
\ee

Using this approximate $\tau$ vs.~$z$ relation and neglecting for the moment the 
contribution from the radiation density, the Friedman equation for $H = H(z)$ 
reduces to
\be
\left({H \over H_{0}}\right)^{2} \approx \Omega_{\rm B}(1 + z)^{3} + 
(1 - \Omega_{\rm B})(1 + z)^{3}{\rm e}^{3\gamma(\tau - \tau_{0})} 
\approx \Omega_{\rm B}(1 + z)^{3} + (1 - \Omega_{\rm B})[\gamma' + 
(1 - \gamma')(1 + z)^{3/2}]^{2}.
\ee  
Notice that for $z = 0$, $H = H_{0}$ and, for $\Omega_{\rm B} = 0$, this result
agrees with that in eq.~12 for $\beta = 0$.  In the high redshift limit, when
$\tau \ll \tau_{0}$ and, including radiation,
\be
({H \over H_{0}})^{2} \rightarrow \Omega_{\rm R}(1 + z)^{4} + [\Omega_{\rm B} + 
(1 - \Omega_{\rm B})(1 - \gamma')^{2}](1 + z)^{3}.
\ee

If the high redshift evolution of this model is to be consistent with that of 
the $\Lambda$CDM concordance model, $\Omega_{\rm B} + (1 - \Omega_{\rm B})(1 - 
\gamma')^{2} = 0.042 + 0.958(1 - \gamma')^{2}\approx \Omega_{\rm M}(\Lambda{\rm 
CDM}) = 0.26$.  This suggest that $\gamma' \approx 0.52$ and $\gamma \approx 
0.51$.  If, instead, it is required that the redshift of equal matter and 
radiation densities agree with the WMAP determined value~\cite{komatsu},
\be
\Omega_{\rm R}(1 + z_{eq}) = \Omega_{\rm B} + (1 - \Omega_{\rm B})(1 - \gamma')^{2},
\ee
we find $\gamma = 0.52 \pm 0.06$.  For $\beta = 0$ and $\gamma = 0.52$, $\tau_{0} 
= 0.97$ and the present age of the Universe in this model is $t_{0} = 13.2$~Gyr.  

As a check on our approximation to the full Friedman equation, where $\rho_{\rm B}$ 
was neglected compared to $\rho_{\rm CDM}$ in order to find the $z - \tau$ relation, 
we note that as $z$ decreases, the ratio of $\rho_{\rm B}$ to $\rho_{\rm CDM}$ 
decreases from $\Omega_{\rm B}/(1 - \Omega_{\rm B}){\rm e}^{-3\gamma\tau_{0}} 
\approx 0.20$ at high $z$, to $\Omega_{\rm B}/(1 - \Omega_{\rm B}) \approx 0.04$ 
at $z = 0$.  From early times to the present, the number of cold dark matter 
particles in a comoving volume increases by a factor of $\sim 4.5$.  Notice 
that for $\beta = 0$, the early ISW effect constraint on $\gamma$ differs 
little without or with baryons, $0.50 \pm 0.06$ versus $0.52 \pm 0.06$, for 
\omb = 0 and \omb = 0.042 respectively.

A discussion of the SNIa constraint in this case ($\beta = 0$) is included 
in the exploration of the more general case ($\beta \neq 0$) considered next.  

\subsection{$\Omega_{\rm B} \neq 0, \gamma \neq 0, \beta \neq 0$}

Based on our analyses of the \omb = 0 case which allowed for non-zero values 
of both $\beta$ and $\gamma$, along with our analysis of the case where \omb 
= 0.042 and either $\beta$ or $\gamma$ = 0, the redshift dependence of the 
full Friedman equation can be approximated as
\be
\left({H \over H_{0}}\right)^{2} \approx \Omega_{\rm R}(1 + z)^{4} + 
\Omega_{\rm B}(1 + z)^{3} + {(1 - \Omega_{\rm B}) \over (1 - \beta)^{2}}\left[\gamma'
+ (1 - \gamma' - \beta)(1 + z)^{3(1 - \beta)/2}\right]^{2}.
\ee
For $z = 0$, $H = H_{0}$, and the age of the Universe is
\be
\tau_{0} \equiv H_{0}t_{0} = {2 \over 3\gamma}{\rm ln}\left({1 - \beta 
\over 1 - \gamma' - \beta}\right).
\ee

\begin{figure}[!ht]
\begin{center}
\epsfig{file=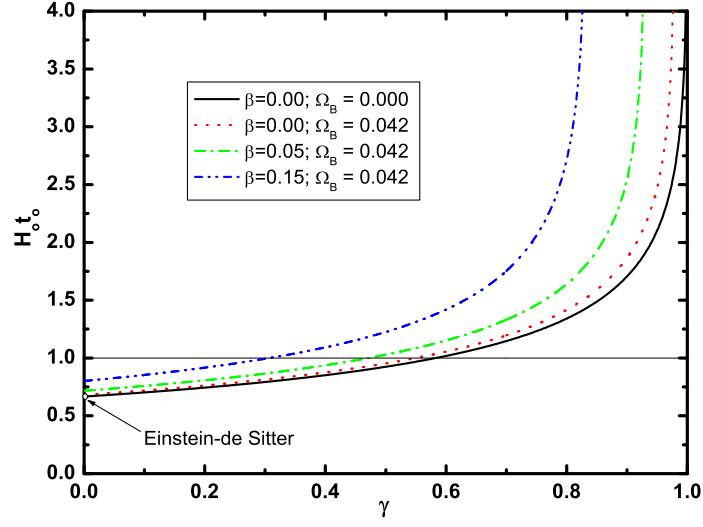, width=4.0truein,height=3.125truein}
\caption{The age of the Universe in units of the Hubble age, 
$\tau_{0} \equiv H_{0}t_{0}$, as a function of $\gamma$ for 
selected values of $\beta$, for models with and without baryons.} 
\label{fig:tau0}
\end{center}
\end{figure}

In Figure~\ref{fig:tau0} are shown the $\tau_{0} - \gamma$ relations for selected values
of \omb and $\beta$.  For $\gamma~\ga 0.45$ and $\beta \geq 0$, $\tau_{0}~\ga 0.9$.  For $\gamma = 0$, $\tau_{0} = 2/3(1 - \Omega_{\rm B})^{-1/2}(1 - \beta)^{-1}$.  Notice that 
as $\gamma \rightarrow 0$ and $\beta \rightarrow 1$, $\tau_{0} \rightarrow \infty$, as 
it should since this case is equivalent to the steady-state model where there is no 
beginning to the Universe.

\begin{figure}[!ht]
\begin{center}
\epsfig{file=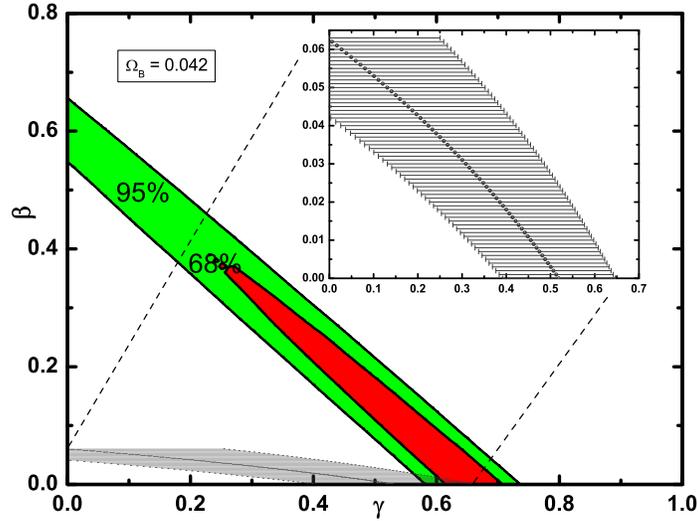, width=4.0truein,height=3.125truein}
\caption{As Figure~\ref{fig:betagamma0}, now including baryons.  The 
68\% (red) and 95\% (green) contours in the $\beta - \gamma$ plane, 
derived for \omb = 0.042 from the SNIa data~\cite{sn1a}, are shown 
along with the 95\% confidence band (black; see the inset which zooms 
in on this band) consistent with the early ISW constraint that $z_{eq} 
= 3141 \pm 157$~\cite{komatsu}.  The SNIa data are best fit at $\beta 
= 0$ and $\gamma = 0.66$.  The solid black curve (see eq.~34), shown 
in more detail in the inset, is for the best fit relation between 
$\beta$ and $\gamma$ consistent with the observed value of $z_{eq}$.}
\label{fig:betagamma042}
\end{center}
\end{figure}

At high redshifts,
\be
\left({H \over H_{0}}\right)^{2} \approx \Omega_{\rm R}(1 + z)^{4} + 
\Omega_{\rm B}(1 + z)^{3} + (1 - \Omega_{\rm B}){\left(1 - \gamma' - \beta 
\over 1 - \beta \right)^{2}}(1 + z)^{3(1 - \beta)}.
\ee
Equating $\rho_{\rm R}$ to $\rho_{\rm M} = \rho_{\rm B} + \rho_{\rm CDM}$ at 
$z = z_{eq}$ leads to a constraint on the $\gamma - \beta$ relation,
\be
\gamma = (1 - \beta)[(1 - \Omega_{\rm B})^{1/2} - (\Omega_{\rm R}(1 + z_{eq}) 
- \Omega_{\rm B})^{1/2}(1 + z_{eq})^{3\beta/2}].
\ee

\begin{figure}[!ht]
\begin{center}
\epsfig{file=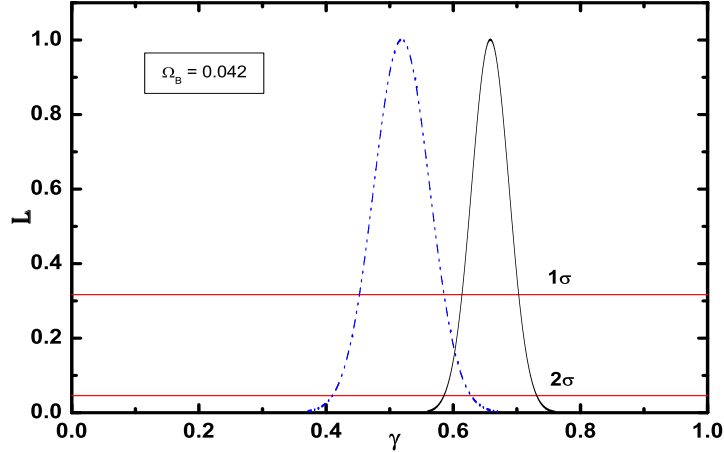, width=4.0truein,height=3.125truein}
\caption{The likelihood functions for $\gamma$ inferred from the SNIa 
data (solid/gray) and from the $z_{eq}$ constraint (dot-dashed/blue) 
for $\beta = 0$ and \omb = 0.042.} 
\label{fig:probgamma22}
\end{center}
\end{figure}
 
In Figure~\ref{fig:betagamma042} are shown the 68\% and 95\% contours in the 
$\beta - \gamma$ plane for \omb = 0.042 from the SNIa data~\cite{sn1a} and 
from the early ISW effect (eq.~34) constrained by the WMAP value of $z_{eq}$ 
\cite{komatsu}.  For the SNIa data~\cite{sn1a} the best fit occurs for $\beta 
= 0$ and $\gamma = 0.66 \pm 0.04$, in some conflict with the best fit to 
$z_{eq}$ for $\beta = 0$, for which $\gamma = 0.52 \pm 0.06$.  As may be seen 
from Figure~\ref{fig:betagamma042}, these two constraints diverge further from 
each other as $\beta$ increases from zero.  In Figure~\ref{fig:probgamma22}, 
are shown the probability distribution functions for $\gamma$ derived from the 
SNIa data~\cite{sn1a} and from $z_{eq}$~\cite{komatsu} for $\beta = 0$ and 
\omb = 0.042.  Note that there is some tension between these two independent 
constraints.  Indeed, if the best fit SNIa values of $\beta$ and $\gamma$ 
are used to predict the redshift of equal matter and radiation densities, 
$(1 + z_{eq})_{\rm SNIa} = 1798^{+536}_{-552}$, which is some $2.5\sigma$ 
away from the WMAP value~\cite{komatsu}.
  
\section{Summary And Conclusions}
\label{concl}

As an alternative to the standard, $\Lambda$CDM model, we have explored here a class 
of models whose late-time acceleration is driven by the creation of cold dark matter.
In constrast to the $\Lambda$CDM model, the models considered here have no cosmological constant or vacuum energy.  While the dark energy models have two or more adjustable parameters (the dark energy equation of state, $w$, and its evolution with redshift, 
in addition to the choice of $\Omega_{\rm DE}$), the flat $\Lambda$CDM model has 
only one free parameter, $\Omega_{\Lambda} = 1 - \Omega_{\rm CDM}$.  Here we have 
considered a class of two-parameter (\{$\beta,\gamma$\}) models. We found that the 
supernovae data prefer $\beta = 0$ and that the early ISW effect is consistent 
with this choice.  For this subset of creation-driven models, with only one free 
parameter ($\gamma$), the high-redshift evolution of the Universe is qualitatively 
indistinguishable from that of the $\Lambda$CDM model, while the recent evolution 
is sufficiently similar to it to allow consistency with the SNIa data.  From the 
high redshift constraint on the WMAP determined value of $z_{eq}$~\cite{komatsu} 
provided by the early ISW effect, we determined $\gamma$(ISW) = 0.52.  For $\beta 
= 0$ and this value for $\gamma$, $H_{0}t_{0} = 0.97$, consistent with the estimate $H_{0}t_{0} = 1.00$ from the $\Lambda$CDM model.  This choice of $\gamma$ corresponds 
to the WMAP determined value of $z_{eq} = 3142$, consistent, within the uncertainties, 
with the $\Lambda$CDM value of $z_{eq} = 3223$.  However, for $\beta = 0$, this choice 
of $\gamma$ provides a poor fit to the SNIa data~\cite{sn1a}.  In contrast, the SNIa 
data prefer $\gamma = 0.66$ which, for $\beta = 0$, corresponds to $H_{0}t_{0} = 
1.13$, in good agreement with the $\Lambda$CDM value and with estimates of the age 
of the Universe.  But, for $\beta = 0$ and $\gamma = 0.66$, the redshift of equal 
matter and radiation densities is $z_{eq} = 1798$, in conflict (at $\sim 2.6\sigma$) 
with the WMAP determined value.

While these models, which are consistent with the SNIa data, offer an intriguing 
alternative to the standard, $\Lambda$CDM concordance model, they face challenges.  
For example, although the one-parameter ($\beta \neq 0$, $\gamma = 0$) model with 
baryons provides a natural solution to the observed, late-time acceleration of the 
Universe, without any need for fine-tuning, its early evolution is baryon-dominated 
and inconsistent with the early ISW effect as constrained by the CMB data.  In 
general, there is a clear tension in this class of models between the SNIa data 
and the independent, high redshift constraint from the observed early ISW effect.  
It should also be noted that in these models the ratio of dark matter (CDM plus 
baryons) to baryons increases during the recent evolution of the Universe from 
$\rho_{\rm M}/\rho_{\rm B} \approx 6.0$ at high redshifts ($z~\ga 10$) to 
$\rho_{\rm M}/\rho_{\rm B} \approx 24$ at present ($z = 0$).  Correspondingly, 
the baryon fraction, $f_{\rm B} \equiv \rho_{\rm B}/\rho_{\rm M}$ decreases from 
$f_{\rm B} \approx 0.17$ at high redshifts to $f_{\rm B} \approx 0.04$ at present.  
While this latter value appears to be in conflict with the x-ray cluster baryon 
fraction~\cite{clusters}, if clusters were formed at sufficiently high redshifts, 
their baryon fraction may be representative of the earlier, higher value of 
$f_{\rm B}$.  Due to the recent increase in the density of CDM, the late-time 
growth of structure and of the cluster baryon fraction in these models will 
differ from that of the $\Lambda$CDM concordance model.  Therefore, before 
ruling out these models, it might be worthwhile to explore their late-time 
evolution more carefully, especially with regard to predictions for the CMB 
and for the growth of large scale structure.

\begin{acknowledgments}
GS acknowledges informative conversations and correspondence with E. Komatsu.
The research of GS is supported at The Ohio State University by a grant from 
the US Department of Energy.  The work reported here was done when GS was a 
Visiting Professor at IAG -- USP and is supported by the  grant from FAPESP.  
RCS is supported by a Fellowship from FAPESP No. 08/52890-1, and JASL is partially supported by CNPq and FAPESP under Grants 304792/2003-9 
and 04/13668-0, respectively. 
\end{acknowledgments}

\end{document}